\documentclass[aps,pre,preprint,groupedaddress]{revtex4}
\usepackage{graphicx}
\usepackage{latexsym}
\usepackage{amsmath,amssymb}

\begin{document}

%\preprint{}
%%\begin{CJK*}{JIS}{}

\title{
Numerical diagonalization analysis of the
criticality of the
$(2+1)$-dimensional $XY$ model:
Off-diagonal Novotny's method
}

\author{Yoshihiro Nishiyama} %% ($B@>;3(B $BM390(B)}
%\thanks{}
\affiliation{Department of Physics, Faculty of Science,
Okayama University, Okayama 700-8530, Japan}

\date{\today}

\begin{abstract}
The criticality of the $(2+1)$-dimensional $XY$ model
is investigated with the numerical diagonalization method.
So far, it has been considered that the diagonalization method
would not be very suitable for analyzing the criticality 
in large dimensions ($d \ge 3$);
in fact,
the tractable system size with the diagonalization method
is severely restricted.
In this paper, we employ Novotny's method,
which enables us to treat
a variety of system sizes
$N=6,8,\dots,20$ ($N$: the number of spins constituting a cluster).
For that purpose,
we develop an off-diagonal version of Novotny's method
to adopt the
off-diagonal (quantum-mechanical $XY$) interaction.
Moreover, in order to improve the finite-size-scaling
behavior,
we tune the coupling-constant parameters to a scale-invariant point.
As a result,
we estimate the critical indices as
$\nu=0.675(20)$ and
$\gamma/\nu=1.97(10)$.
%The estimates are accordant with
%the existing values for the $d=3$ $XY$ universality class.
\end{abstract}

%\keywords{}

\maketitle

%%\end{CJK*}

\section{\label{section1}Introduction}

It has been considered that the diagonalization
method would not be
very suitable for analyzing the criticality
in large dimensions $d \ge 3$.
In fact, as the system size enlarges,
the number of spins constituting a cluster
increases rapidly in $d \ge 3$,
and the dimensionality of the Hilbert space soon exceeds the
limitation of
available computer resources.
Such a severe limitation as to the tractable system size
prevents us from making a systematic analysis
of the simulation data.

To cope with this difficulty, 
Novotny proposed a transfer-matrix formalism
\cite{Novotny90,Novotny92,Novotny91},
which enables us to
construct a transfer-matrix unit with an arbitrary 
(integral) number of spins $N$; 
note that conventionally, the number of spins 
is restricted within $N=2^{d-1},3^{d-1},\dots$.
As a demonstration,
Novotny simulated the Ising model in $d \le 7$
systematically \cite{Novotny91}.
Meanwhile, 
it has been shown that the idea is applicable to
a wide class of
systems 
such as
the frustrated Ising model
\cite{Nishiyama04}
and
the quantum-mechanical Ising model under the transverse magnetic field
\cite{Nishiyama07}.

In this paper, we extend the Novotny method to adopt
the off-diagonal (quantum-mechanical $XY$) interaction;
see the Hamiltonian, Eq. (\ref{Hamiltonian}),
mentioned afterward.
Actually, as mentioned above,
the use of Novotny's method has been restricted
within the case of the diagonal (Ising-type) interaction.
As a demonstration, we apply the method to the $(2+1)$-dimensional $XY$ model
with a variety of system sizes
$N=6,8,\dots,20$.
Taking the advantage that a series of system sizes are available,
we made a systematic finite-size-scaling analysis of the simulation data.
As a result,
we estimate the critical indices 
as $\nu=0.675(20)$ and $\gamma/\nu=1.97(10)$.
Recent developments on the $d=3$ $XY$ universality class
are overviewed in Ref. \cite{Barmatz07}
with an emphasis on the microgravity-environment experiment;
see also Refs. \cite{Buteral97,Guida98,Kleinert05,Burovski06,Campostrini06}.
Our method provides an alternative approach
to the $d=3$ $XY$ universality class.

To be specific, we consider the following Hamiltonian
for the $(2+1)$-dimensional $XY$ model 
\cite{Costa03,Leonel06,Dely06}
with the extended interactions
\begin{equation}
\label{Hamiltonian}
{\cal H}=
-J_{NN} \sum_{\langle ij \rangle} (S^x_i S^x_j+S^y_i S^y_j)
-J_{NNN} \sum_{\langle\langle ij \rangle\rangle} (S^x_iS^x_j+S^y_iS^y_j)
+D_{\Box} \sum_{[ijkl]} (S^z_i+S^z_j+S^z_k+S^z_l)^2
+D \sum_i (S^z_i)^2
   .
\end{equation}
Here,
the quantum spin-1 ($S=1$) operators $\{ {\bf S}_i \}$
are
placed at each square-lattice point $i$.
The summations,
$\sum_{\langle ij\rangle}$, 
$\sum_{\langle\langle ij\rangle\rangle}$,
and 
$\sum_{[ijkl]}$,
run over all possible
nearest-neighbor, next-nearest-neighbor,
and plaquette spins,
respectively.
The parameters,
$J_{NN}$, $J_{NNN}$, and $D_\Box$, are the corresponding coupling constants.
The single-ion anisotropy $D$,
drives the system from the $XY$ phase
($D < D_c$) to the large-$D$ phase 
($D > D_c$).
(In the large-$D$ phase, the ground state is magnetically disordered,
accompanied with
a finite excitation gap.)
Our aim is to survey the criticality 
by means of
the off-diagonal Novotny method.
%The finite-temperature phase transition
%of the model
%was studied extensively with the 
%schuwinger boson mean field theory ??.

The Hamiltonian (\ref{Hamiltonian}) has a number of 
tunable parameters.
We fixed them to
\begin{equation}
\label{FP_intro}
(J_{NN},J_{NNN},D_\Box)=(
 0.158242810160,
0.058561393564,
0.10035104389)  ,
\end{equation}
and survey the $D$-driven phase transition.
As explicated in the Appendix,
around the point (\ref{FP_intro}),
the finite-size-scaling behavior improves significantly;
the irrelevant interactions cancel out,
because the point (\ref{FP_intro})
is a scale-invariant point with respect to the real-space decimation
shown in Fig. \ref{figure7}.
Such an elimination of finite-size corrections 
has been utilized successfully
to analyze the criticality
of the classical systems such
as the Ising model \cite{Blote96,Nishiyama06} and the lattice $\phi^4$ theory
\cite{Ballesteros98,Hasenbusch99}.
We adopt the idea to investigate a quantum-mechanical system, Eq. (\ref{Hamiltonian}).

The rest of this paper is organized as follows.
In Sec. \ref{section2},
we develop an off-diagonal version of the Novotny method.
In Sec. \ref{section3}, employing this method,
we simulate the $(2+1)$-dimensional
$XY$ model (\ref{Hamiltonian}).
In Sec. \ref{section4}, we present the summary and discussions.
In the Appendix, we determine a scale-invariant point
(\ref{FP_intro}) with respect to the real-space decimation
shown in
Fig. \ref{figure7}.

\section{\label{section2}Off-diagonal Novotny's method}

In this section,
we explain the simulation scheme.
As mentioned in the Introduction,
we develop an off-diagonal version of Novotny's method
to simulate the $(2+1)$-dimensional $XY$ model (\ref{Hamiltonian});
so far, the Novotny method has been applied to the 
case of Ising-type interactions 
\cite{Novotny90,Novotny92,Novotny91,Nishiyama04,Nishiyama07}.

\begin{figure}
\includegraphics[width=100mm]{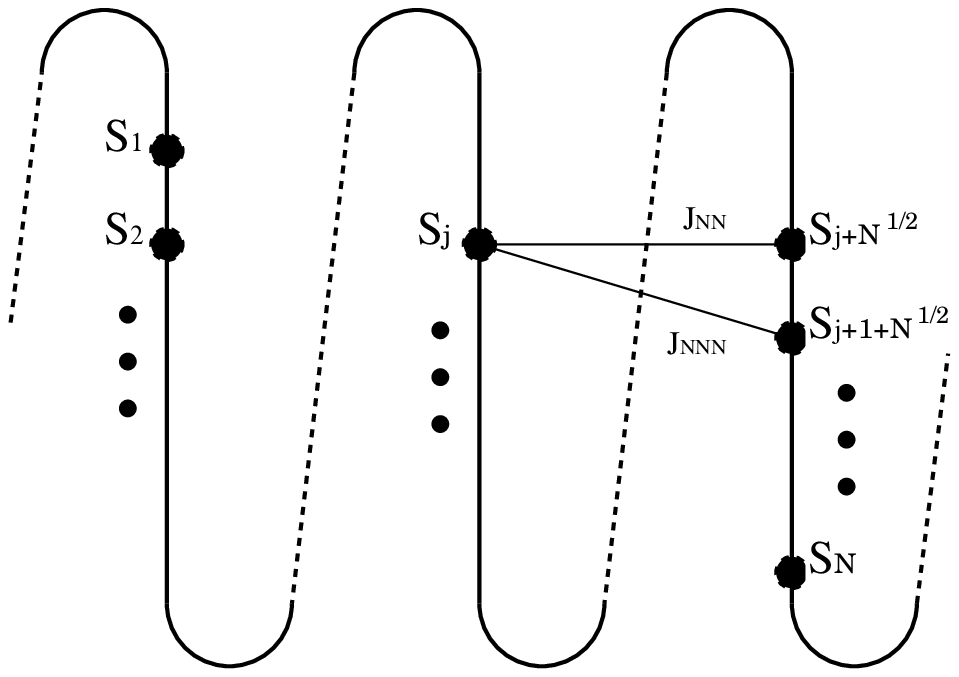}%
\caption{
\label{figure1}
A schematic drawing of the spin cluster for the
$d=2$ quantum $XY$ model (\ref{Hamiltonian}) is presented.
As indicated above, the spins constitute a 
one-dimensional alignment $\{ S_i \}$ ($i=1,2,\dots,N$),
and the dimensionality is lifted to $d=2$ by the bridges over the 
$(N^{1/2})$th-neighbor pairs.
This is a basic idea of Novotny's diagonalization method.
We need to develop an off-diagonal version of Novotny's method,
because we have to adopt the quantum $XY$ interaction 
(\ref{Hamiltonian});
see Sec. \ref{section2}.
}
\end{figure}

To begin with, we explain the basic idea of Novotny's method.
Novotny's method allows us to 
construct a cluster with
an arbitrary number of spins;
see Fig. \ref{figure1}.
As indicated, 
the basic structure of the cluster is
one-dimensional.
The dimensionality is lifted to $d=2$ by the bridges over the $(\sqrt{N})$-th-neighbor
pairs.
Because the basic structure is one-dimensional,
we are able to construct a cluster with an arbitrary 
(integral) number of spins $\{ {\bf S}_i \}$ ($i=1,2,\dots,N$);
note that naively, the number of spins is restricted within
$N=4,9,16,\dots$ in $d=2$.

We formulate the above idea explicitly.
We propose the following expression 
\begin{equation}
\label{Hamiltonian_Novotny}
{\cal H}=
  - J_{NN} ( H_{XY} (1) + H_{XY} (\sqrt{N}) )
- J_{NNN} ( H_{XY}(\sqrt{N}+1)+H(\sqrt{N}-1) )
  + D_\Box  H_\Box (\sqrt{N})
 +  D \sum_{i=1}^N (S_i^z)^2    ,
\end{equation}
for the Hamiltonian of the $(2+1)$-dimensional $XY$ model (\ref{Hamiltonian}).
The component $H_{XY(\Box)}(v)$ describes the 
$XY$ (plaquette) interaction over the $v$-th neighbor pairs;
see Fig. \ref{figure1}.
Because the quantum $XY$ interaction,
$H_{XY}(v)$, is an off-diagonal one,
we need to develop an off-diagonal version
of Novotny's method.
We propose the following expression
\begin{equation}
\label{H_XY}
H_{XY}(v)=\sum_{i=1}^N (P^v S^x_i P^{-v} S^x_i +P^v S^y_i P^{-v} S^y_i)       .
\end{equation}
This formula serves as a basis of
the off-diagonal Novotny method.
The symbol $P$
denotes the translation operator by one lattice spacing;
\begin{equation}
P |  S_1,S_2,\dots,S_N\rangle = |S_N,S_1,\dots,S_{N-1}\rangle   .
\end{equation}
(We impose the periodic boundary condition,
$S_{N+1}=S_1$.)
Here, the base
$| \{ S_k \} \rangle $ diagonalizes the
$\{ S^z_i \}$ operators; namely, it satisfies
\begin{equation}
S^z_l |\{S_k\}\rangle = S_l |\{ S_k\}\rangle   ,
\end{equation}
for each $l=1,2,\dots,N$.
The insertions of the operators $P^{\pm v}$ 
in Eq. (\ref{H_XY}) introduce 
the $v$-th neighbor interaction along the alignment of spins $\{ S_i \}$;
symbolically, the operator $P^v S^\alpha_i P^{-v}$ may be written as
$S^\alpha_{i+v}$.
On the other hand, as for $H_\Box(v)$,
we adopt the conventional idea based on the diagonal Novotny method \cite{Novotny90}.
That is, its diagonal elements 
$\{ \langle \{S_k\}|H_\Box(v)|\{S_k\}\rangle  \} $ 
are given by
\begin{equation}
\label{H_Box}
\langle \{ S_k\} | H_{\Box}(v) | \{ S_k\} \rangle =
\langle \{S_k \} | P^v  T | \{S_k\} \rangle
     ,
\end{equation}
with the four-spin interaction
\begin{equation}
\langle \{S_k\} | T | \{T_k\}\rangle
     = \sum_{l=1}^N S_lS_{l+1}T_lT_{l+1}   .
\end{equation}
Similarly, the insertion of $P^v$ introduces the $v$-th neighbor interaction.
However, in the diagonal scheme (\ref{H_Box}),
one operation of $P^v$ suffices;
note that in the off-diagonal formalism (\ref{H_XY}),
two operations $P^{\pm v}$ are required.
Because
each operation $P^{\pm v}$ requires huge computational effort,
the off-diagonal scheme 
is computationally demanding.
Afterward, we provide a number of formulae useful
for the practical implementation of the algorithm.

The above formulae complete the formal basis of our simulation scheme.
Aiming to improve the simulation result,
we implement the following symmetrization technique \cite{Novotny92}.
That is, we symmetrize
the component $H_{XY,\Box}(v)$ by replacing it with
\begin{equation}
H_{XY,\Box}(v) \to (H_{XY,\Box}(v)+H_{XY,\Box}(-v))/2
   .
\end{equation}
This replacement restores the symmetry between
the ascending, $S_1,S_2,\dots,S_N$, and the descending,
$S_N,S_{N-1},\dots,S_1$, directions completely.

Last, we provide a number of formulae that may be useful
in the practical implementation of the algorithm.
We utilize the
translationally invariant bases $\{ |k,n\rangle  \}$,
which diagonalize the operator $P$;
\begin{equation}
P |k,n \rangle  = e^{ik} |k,n \rangle .
\end{equation}
Here, the wave number $k$ runs over a
Brillouin zone
$k=2\pi M /N$ ($M$: integer),
and
the index $n$ specifies the state within the subspace $k$.
As anticipated, the bases $ \{ |k,n\rangle \} $
are useful to obtain an explicit representation of the formulae
mentioned above.
For instance, the first term of 
the formula (\ref{H_XY}) is represented by
\begin{equation}
\langle k,n   | \sum_{j=1}^N S^x_j  P^v  S^x_j P^{-v}  | k,m \rangle  
  =  \sum_{j=1}^N \sum_{k',n'} 
   \langle k,n|S^x_j| k',n'\rangle\langle k',n' | S^x_j | k,m \rangle e^{i(k'-k)v}  
  ,
\end{equation}
in terms of the frame $\{ |k,n\rangle \}$.
Because the parameter $v$ is, in general, an irrational number,
the oscillating factor $e^{i(k'-k)v}$ is incommensurate 
with respect to the lattice periodicity.
Hence,
the intermediate summation $\sum_{k'}$ has to be treated carefully;
namely,
each Brillouin zone $\{ k' \}$ is no longer equivalent.
We accepted the following symmetrized sum
\begin{equation}
\sum_{k'} a_{k'} = 
\frac{1}{2}a_{-\pi}+a_{   -\pi +2\pi/N}+a_{-\pi +4\pi/N}
+\dots+a_{\pi-2\pi/N}+\frac{1}{2}a_{\pi}
   ,
\end{equation}
with respect to a summand $a_{k'}$.
Here,
the denominators of the first and the last
terms
compensate the duplicated sum at the edges of the Brillouin zone
$[-\pi,\pi]$.
(Similarly, we obtain an explicit representation for 
$H_\Box(v)$
via the conventional Novotny method
\cite{Novotny90,Novotny92}.)
Provided that the explicit matrix elements of $H_{XY,\Box}(v)$ are 
at hand,
we are able to perform the numerical diagonalization of 
the Hamiltonian (\ref{Hamiltonian_Novotny}).
The results are shown in the next section.

Last, we make an overview of the $S=1$ $XY$ model (\ref{Hamiltonian}).
As mentioned in the Introduction, the model has been studied in Refs. 
\cite{Costa03,Leonel06,Dely06}.
In the case of $d=1$ dimension,
the criticality ($D$-driven phase transition) was investigated in detail
\cite{Botet83}.
According to Ref. \cite{Botet83},
for sufficiently large $D$,
a magnetically disordered ground state (large-$D$ phase) appears,
and the criticality is identical to that of the
classical counterpart
in $d+1(=2)$ dimensions (KT transition).
Unfortunately, a naive extension to the case of $S=1/2$ 
is not appropriate,
because
the $D$ anisotropy, $D (S^z_i)^2$, reduces to a constant term, $D/4$.
(Moreover, the transverse magnetic field violates the $XY$ symmetry, and 
the criticality changes into the Ising type.)
As a matter of fact,
it is difficult to realize a ground-state phase transition
for the $S=1/2$ model
without violating the translational 
invariance and the rotational symmetry.
(Possibly, the double-plane $S=1/2$ model may exhibit a desirable criticality
by tuning 
the inter-plane interaction. However, this model is too complicated.)
Hence, we consider the $S=1$ $XY$ model (\ref{Hamiltonian})
with the $D$-anisotropy term.

\section{\label{section3}Numerical results}

In this section, we analyze the criticality of the
$(2+1)$-dimensional $XY$ model, Eq. (\ref{Hamiltonian}).
As mentioned in the Introduction,
the coupling-constant parameters $(J_{NN},J_{NNN},D_\Box)$
are set to the scale-invariant point (\ref{FP_intro}).
Thereby,
we survey the
$D$-driven phase transition with the finite-size scaling.
In order to diagonalize the Hamiltonian,
we utilize
the off-diagonal Novotny method developed in Sec. \ref{section2}.
Owing to this method, we
treat a variety of system sizes $N=6,8,\dots,20$.
The linear dimension $L$ of the cluster is given by
\begin{equation}
L=\sqrt{N}        ,
\end{equation}
because the $N$ spins constitute a rectangular cluster;
see Fig. \ref{figure1}.

\subsection{\label{section3_1}Transition point}

In this section, we provide an evidence of the
$D$-driven phase transition,
and estimate the critical point
$D_c$ with the finite-size scaling.

In Fig. \ref{figure2},
we plot the scaled energy gap $L \Delta E$ for various
$D$, and $N=6,8,\dots,20$ with the other coupling constants fixed to Eq. (\ref{FP_intro}).
The symbol $\Delta E$ denotes the first-excitation gap.
According to the finite-size scaling,
the scaled energy gap $L \Delta E$ should be invariant
at the critical point.
Indeed,
we observe an onset of the $D$-driven phase transition
around $D \approx 1$.

In Fig. \ref{figure3},
we plot the approximate transition point $D_c(L_1,L_2)$
for $(2/(L_1+L_2))^3$ with $6 \le N_1 < N_2 \le 20$ and $L_{1,2}=\sqrt{N_{1,2}}$;
the validity of the $1/L^3$-extrapolation scheme (abscissa scale)
is considered at the end of this section.
Here,
the approximate transition point 
$D_c(L_1,L_2)$ denotes a scale-invariant point
with respect to a pair of system sizes $(L_1,L_2)$.
Namely, the approximate transition point satisfies
the equation
\begin{equation}
\label{TP_def}
L_1 \Delta E(L_1)|_{D=D_c(L_1,L_2)} = L_2 \Delta E(L_2)|_{D=D_c(L_1,L_2)}
   .
\end{equation}
The least-squares fit to the data of Fig. \ref{figure3} yields
$D_c=0.9569(83)$
in the thermodynamic limit, $L \to \infty$.
As a reference, 
we calculated $D_c=0.9744(68)$ through the $1/L^4$-extrapolation scheme.
Considering the deviation as an error indicator,
we estimate the critical point as
\begin{equation}
\label{transition_point}
D_c=0.957(25)
  .
\end{equation}

Let us mention a few remarks.
First, we consider the abscissa scale 
 $1/L^3$ utilized in Fig. \ref{figure3}.
Naively, the scaling theory predicts that 
dominant corrections to $D_c$ should scale like
$1/L^{\omega+1/\nu}$ with $\omega=0.785(20)$ and $\nu=0.6717(1)$ 
\cite{Campostrini06}.
On one hand, 
in our simulation, such dominant corrections should be suppressed
by tuning the coupling constants to Eq. (\ref{FP_intro});
see the Appendix.
The convergence 
to the thermodynamic limit
may be 
accelerated 
\cite{Nishiyama06}.
(For extremely large system sizes,
the singularity $1/L^{\omega+1/\nu}$ may emerge.)
Hence, in Fig. \ref{figure3}, we set the abscissa scale to $1/L^3$.
Second, 
we argue a consistency between the finite-size scaling
and the real-space decimation;
in the Appendix, we made a fixed-point analysis (\ref{FP_app}),
regarding $D$ as a unit of energy $D=1$ (\ref{Dwo1ni}).
This proposition $D=1$ is quite consistent with
the above scaling result
(\ref{transition_point}),
validating 
the fixed-point analysis in the Appendix.
In other words, around the fixed point (\ref{FP_app}),
corrections to scaling may cancel out satisfactorily.
Encouraged by this consistency,
in Sec. \ref{section3_3},
we survey the criticality rather in detail.

\subsection{Comparison with the conventional $XY$ model}

In the preceding section, we simulated the $XY$ model (\ref{Hamiltonian})
with the finely tuned coupling constants (\ref{FP_intro}).
As a comparison, in this section,
we provide the data for the conventional $XY$ model.
That is, 
we turn off the extended coupling constants,
setting the interactions to
$(J_{NN},J_{NNN},D_\Box)=(0.2,0,0)$ tentatively.

In Fig. \ref{figure4},
we plot the scaled energy gap 
$L\Delta E$
for various $D$ and $N=6,8,\dots,20$.
We observe an onset of the $D$-driven phase transition
around $D \approx 1.1$.
However, the data are scattered, as compared to those of Fig. \ref{figure2}.
In fact, in Fig. \ref{figure4},
the location of the transition point appears to be less clear.
This result demonstrates that
the finely-tuned coupling constants (\ref{FP_intro})
lead to elimination of finite-size corrections.

\subsection{\label{section3_3}Critical exponents}

In Sec. \ref{section3_1}, we observe an onset of the 
$D$-driven phase transition.
In this section, we calculate the critical exponents,
$\nu$ and $\gamma/\nu$, based on the finite-size scaling.

In Fig. \ref{figure5},
we plot the approximate critical exponent
\begin{equation}
\label{nu_def}
\nu(L_1,L_2)=
\frac{
 \ln(L_1/L_2)    }{
     \ln ( \partial_D (L_1 \Delta E(L_1))
   /\partial_D (L_2 \Delta E(L_2)))|_{D=D_c}}  %  (L_1,L_2)}   }
         ,
\end{equation}
for $(2/(L_1+L_2))^2$ with 
$6 \le N_1 < N_2 \le 20$ ($L_{1,2}=\sqrt{N_{1,2}}$), and
$D_c=0.957$ [Eq. (\ref{transition_point})];
afterward, we consider the abscissa scale, $1/L^2$.
The least-squares fit to these data yields
$\nu=0.675(16)$.
As a reference,
we calculated $\nu=0.687(11)$ through the $1/L^3$-extrapolation scheme.
Considering the deviation as an error indicator,
we estimate the critical exponent as
\begin{equation}
\label{kotae1}
\nu=0.675(20)
  .
\end{equation}

In Fig. \ref{figure6},
we plot the approximate critical exponent
\begin{equation}
\label{gwn_def}
\gamma/\nu=   
    \ln(\chi_\perp(L_1)/\chi_\perp(L_2))|_{D=D_c}    %  (L_1,L_2)}  
/\ln(L_1/L_2)
   ,
\end{equation}
for $(2/(L_1+L_2))^2$ with
$6 \le N_1 < N_2 \le 20$ ($L_{1,2}=\sqrt{N_{1,2}}$), and
$D_c=0.957$ [Eq. (\ref{transition_point})].
Here, the transverse susceptibility $\chi_\perp$ is given by the resolvent form
\begin{equation}
\label{transverse_sus}
\chi_\perp=\frac{1}{N}
  \langle g |  M_x \frac{1}{{\cal H}-E_g}M_x | g \rangle
    ,
\end{equation}
with the ground state $|g\rangle$ and the ground-state energy $E_g$.
The magnetization $M_x$ is given by $M_x=\sum_{i=1}^N S_i^x$.
The resolvent form (\ref{transverse_sus})
is readily calculated with use of
the continued-fraction method
\cite{Gagliano87}.

The least-squares fit to the data in Fig. \ref{figure6} yields $\gamma/\nu=1.965(61)$.
As a reference, we calculated $\gamma/\nu=2.020(42)$ through the $1/L^3$-extrapolation scheme.
Considering the deviation as an error indicator,
we estimate the critical exponent as
\begin{equation}
\label{kotae2}
\gamma/\nu=1.97(10)
   .
\end{equation}

Last, we argue the abscissa scale $1/L^2$ utilized in Figs. \ref{figure5}
and \ref{figure6}.
Naively, 
the scaling theory predicts that dominant corrections to the critical indices should scale
like $1/L^\omega$
with $\omega=0.785(20)$
\cite{Campostrini06}.
On one hand,
as argued in Sec. \ref{section3_1},
such dominant corrections should be
suppressed by adjusting the coupling constants to Eq. (\ref{FP_intro}),
and the
convergence is accelerated than the naively expected one
\cite{Nishiyama06}.
Hence, we set the abscissa scale to $1/L^2$ in Figs.
\ref{figure5} and \ref{figure6}.

\subsection{Refined data analysis}

In this section, we make an alternative analysis
of the criticality to demonstrate
a reliability of our scheme.

In Figs. \ref{figure_new1} and \ref{figure_new2},
we plot the critical exponents
\begin{equation}
\label{nu_def2}
\nu   %  (L_1,L_2)
= \frac{
 \ln(L_1/L_2)    }{
     \ln ( \partial_D (L_1 \Delta E(L_1))
   /\partial_D (L_2 \Delta E(L_2)))|_{D=D_c(L_1,L_2)}}  %  (L_1,L_2)}   }
         ,
\end{equation}
and
\begin{equation}
\label{gwn_def2}
\gamma/\nu=   
    \ln(\chi_\perp(L_1)/\chi_\perp(L_2))|_{D=D_c(L_1,L_2)}    %  (L_1,L_2)}  
/\ln(L_1/L_2)
   ,
\end{equation}
respectively,
for $(2/(L_1+L_2))^2$ with
$6 \le N_1 < N_2 \le 20$.
Here, these exponents are calculated at the
approximate critical point
$D=D_c(L_1,L_2)$ (\ref{TP_def})
rather than at $D_c=0.957$ as in the preceding section.

Clearly, 
these data, Figs. \ref{figure_new1} and \ref{figure_new2}, 
exhibit accelerated convergence
to the thermodynamic limit
as compared to those of Figs.
\ref{figure5} and \ref{figure6}.
In fact, the least-squares fit to these data yields
the estimates $\nu=0.658(5)$ and $\gamma/\nu=1.946(4)$
with suppressed error margins.
Actually,
in Fig. \ref{figure_new2},
the systematic error dominates the insystematic one.
In such a case, one has to
make a detailed consideration of the nature of
corrections to scaling to ensure the accuracy (amount of error margin)
of the extrapolation.
Here, we do not commence making such a consideration,
and accept the estimates, Eqs.
(\ref{kotae1}) and (\ref{kotae2}),
obtained
less ambiguously
in the preceding section.
It is not the purpose of this paper 
to obtain fully
refined estimates for the critical indices.
Such a detailed analysis will be
pursued in the succeeding works.
In fact, the diagonalization method has a potential applicability
to the frustrated magnetism, for which the quantum Monte Carlo
method suffers from the 
notorious sign problem.
The Novotny method would be particularly of use
to explore such a problem.
Actually, in the case of the Ising-type anisotropy,
the Novotny method was applied \cite{Nishiyama07b}
to clarifying the nature of the frustration-driven transition
(Lifshitz point).
The present scheme may provide a basis 
for surveying such a quantum frustrated system with the $XY$-type 
symmetry.

\section{\label{section4}Summary and discussions}

The criticality of the $(2+1)$-dimensional $XY$ model 
(\ref{Hamiltonian})
was investigated with the numerical-diagonalization method.
For that purpose,
we developed an off-diagonal version
of Novotny's diagonalization method (Sec. \ref{section2}),
which enables us to
treat a variety of system sizes
$N=6,8,\dots,20$ ($N$: the number of spins within a cluster).
Moreover, we improved the finite-size-scaling behavior
by adjusting the coupling-constant parameters
to a scale-invariant point (\ref{FP_intro}).

Owing to these improvements,
we could analyze the simulation data systematically
with the finite-size scaling.
As a result,
we estimated the critical indices as
$\nu=0.675(20)$ and $\gamma/\nu=1.97(10)$.
These indices immediately yield
the following critical exponents
\begin{equation}
\label{final_results}
\alpha=-0.025(60)   , \beta=0.348(49)     , \ and \ \gamma=1.330(78)  ,
\end{equation}
through the scaling relations.

Recent developments
on the $d=3$ $XY$ universality class
are overviewed in Ref. \cite{Barmatz07}.
Our diagonalization result (\ref{final_results})
is accordant with
a Monte Carlo result,
$\alpha=-0.0151(3)$, $\beta=0.3486(1)$, and $\gamma=1.3178(2)$  \cite{Campostrini06},
and
a field-theoretical result,
$\alpha=-0.011(4)$, $\beta=0.3470(16)$, and $\gamma=1.3169(20)$
\cite{Guida98}.
(In Ref. \cite{Campostrini06}, 
information from a series-expansion result
is also taken into account.)
%Concerning the amount of errors,
%the present result is almost comparable
%to that of the series expansion, 
%for instance,
%$\alpha$.... \cite{ll}.
To the best of our knowledge,
no numerical-diagonalization result has been reported 
as for the $d=3$ $XY$ universality class.
According to Ref. \cite{Barmatz07},
there arose a discrepancy between
the Monte Carlo simulation and the microgravity-environment experiment;
see also Ref.
\cite{Pogorelov07}.
As a matter of fact,
the microgravity experiment \cite{Lipa03}
reports a critical exponent $\alpha=-0.0127(3)$.
In order to resolve this discrepancy, 
an alternative
scheme
other than the Monte Carlo and series-expansion methods
would be desirable.
Refinement of the diagonalization scheme
through considering the singularity of corrections to scaling
might be significant in order to settle this longstanding issue.

%and it will be explored in the future study.

%The series-expansion and numerical-diagonalization methods
%may be promising candidates
%to resolve this longstanding issue.

\begin{acknowledgments}
This work was supported by a Grant-in-Aid 
(No. 18740234) from Monbu-Kagakusho, Japan.
\end{acknowledgments}

\appendix

\section{Search for a scale-invariant point:
Elimination of finite-size corrections}

As mentioned in the Introduction,
we simulated
the quantum $XY$ model (\ref{Hamiltonian}),
setting
the coupling constants to Eq.
(\ref{FP_intro});
around this point, we observe eliminated
finite-size corrections.
In this Appendix, we explicate the scheme to determine the point
(\ref{FP_intro}).

To begin with,
we explain the technique to suppress the finite-size corrections.
According to Refs. \cite{Blote96,Nishiyama06,Ballesteros98,Hasenbusch99},
the finite-size behavior improves
around the renormalization-group fixed point.
That is,
the irrelevant interactions may cancel out
around the fixed point.
Clearly,
such an improvement of the finite-size behavior
admits us to make a systematic finite-size-scaling analysis
of the simulation data.
To avoid confusion, we stress that the fixed-point analysis
is simply a preliminary one,
and subsequently, we perform large-scale computer simulation
to estimate
the critical exponents.
In this sense,
as for the Monte Carlo simulation,
it might be more rewarding to enlarge the system size
rather than to extend the coupling-constant parameters and adjust them.
On one hand, 
it is significant 
for the numerical diagonalization
to eliminate corrections to scaling,
because its tractable system size
is restricted intrinsically.

In Fig. \ref{figure7},
we present a 
schematic drawing of 
the real-space-decimation procedure.
As indicated,
we consider a couple of rectangular clusters with the 
sizes $2 \times 2$ and $4\times 4$.
These clusters are labeled by the symbols $S$ and $L$, respectively.
Decimating out the spin variables indicated by the symbol $\bullet$ 
within
the $L$ cluster,
we obtain a coarse-grained lattice 
identical to the $S$ cluster.
Our aim is to search for a scale-invariant point
with respect to the real-space decimation.

Before going into the fixed-point analysis,
we set up the simulation scheme for the clusters, $S$ and $L$.
We cast
the Hamiltonian (\ref{Hamiltonian}) into the following plaquette-based expression
\begin{equation}
{\cal H} =  \sum_{[ijkl]} {\cal H}^\Box_{ijkl}    % (J_{NN},J_{NNN},D_\Box)
+D \sum_i (S^z_i)^2  ,
\end{equation}
with the plaquette interaction
\begin{eqnarray}
{\cal H}^\Box_{ijkl} %   (J_{NN},J_{NNN},D_\Box)  &=&
 &=& -\frac{J_{NN}}{2}(
S^x_iS^x_j+S^y_iS^y_j+
S^x_jS^x_l+S^y_jS^y_l+
S^x_kS^x_l+S^y_kS^y_l+
S^x_iS^x_k+S^y_iS^y_k)    \\
     &  &  -
J_{NNN}(
S^x_iS^x_l+S^y_iS^y_l+
S^x_jS^x_k+S^y_jS^y_k)   \\
& & +
D_\Box(S^z_i+S^z_j+S^z_k+S^z_l)^2
         .
\end{eqnarray}
(The denominator of the coefficient $J_{NN}$ compensates the duplicated sum.)
Hence,
the Hamiltonian for the $S$ cluster is given by
\begin{equation}
{\cal H}_S  = {\cal H}^\Box_{1234}    %  [(1+b)J_{NN},(1+b)J_{NNN},D_\Box]
+D \sum_{i=1}^4 (S^z_i)^2    ,
\end{equation}
with the replacement
\begin{equation}
J_{NN,NNN} \to (1+b) J_{NN,NNN}   .
\end{equation}
Here, the parameter $b$ controls the boundary interaction strength,
and hereafter, we set $b=0.7$; we consider the validity of this choice afterward.
The boundary-interaction parameter $b$
interpolates smoothly the open, $b=0$, and periodic, $b=1$,
boundary conditions.
The point is that for the two-site ($L=2$) system,
the bulk interaction, $S^\alpha_1S^\alpha_2$,
and the boundary interaction $S^\alpha_2S^\alpha_1$ coincide each other.
Hence, for the $S$ cluster,
the boundary interaction $b$ is freely tunable
without violating the translation invariance.
We make use of this redundancy to obtain the fixed point reliably.
On the other hand,
the $L$ cluster does not have such
a redundancy, and
the Hamiltonian ${\cal H}_L$ is 
given by Eq. (\ref{Hamiltonian}) with $L=4$ unambiguously.
We diagonalize these Hamiltonian matrices
${\cal H}_{S,L}$ numerically;
note that we employ the conventional diagonalization method,
rather than the off-diagonal Novotny method developed in Sec. \ref{section2}.

With use of the simulation technique developed above,
we search for the fixed point of the real-space decimation.
%%Fig. \ref{figure7}.
We survey the 
parameter space
$(J_{NN},J_{NNN},D_\Box)$,
regarding $D$ as a unit of energy; namely, we set
\begin{equation}
\label{Dwo1ni}
D=1   ,
\end{equation}
throughout this section.
Thereby,
we impose the following conditions
\begin{eqnarray}
\label{SC1}
2 \Delta E_S &=& 4 \Delta E_L    \\
\label{SC2}
\langle S^x_1 S^x_2 \rangle_S &=& \langle \tilde{S}^x_1 \tilde{S}^x_2 \rangle_L \\
\label{SC3}
\langle S^x_1 S^x_4 \rangle_S &=& \langle \tilde{S}^x_1 \tilde{S}^x_4 \rangle_L
   ,
\end{eqnarray}
as a scale-invariance criterion.
The symbol $\Delta E_{S,L}$ denotes the excitation gap for the respective clusters.
The arrangement of the spin variables, $S_{1,2,3,4}^\alpha$ and $\tilde{S}_{1,2,3,4}^\alpha$,
is shown in Fig. \ref{figure7}.
The symbol $\langle \dots \rangle_{S,L}$ denotes the 
ground-state average
for the respective clusters.
The first equality (\ref{SC1}) comes from the scale invariance of 
the scaled energy gap, $L \Delta E$.
(We refer readers to Ref. \cite{Itakura03}, where
the author utilizes
such a critical-amplitude relation 
successfully to analyze the renormalization-group flow numerically.)
On one hand,
the remaining equations, (\ref{SC2}) and (\ref{SC3}),
are the scale-invariance conditions \cite{Swendsen82} 
regarding the correlation functions for the
edge and diagonal spins, respectively.

The conditions, Eqs. (\ref{SC1})-(\ref{SC3}), are the
nonlinear equations 
with respect to $(J_{NN},J_{NNN},D_\Box)$.
In order to obtain the solution,
we employed the 
Newton method, and
found that the following nontrivial solution does exist;
\begin{equation}
\label{FP_app}
(J_{NN},J_{NNN},D_\Box)=(
 0.158242810160,
0.058561393564,
0.10035104389)  
   .
\end{equation}
The last digits may be uncertain because of the round-off errors.

\begin{figure}
\includegraphics[width=100mm]{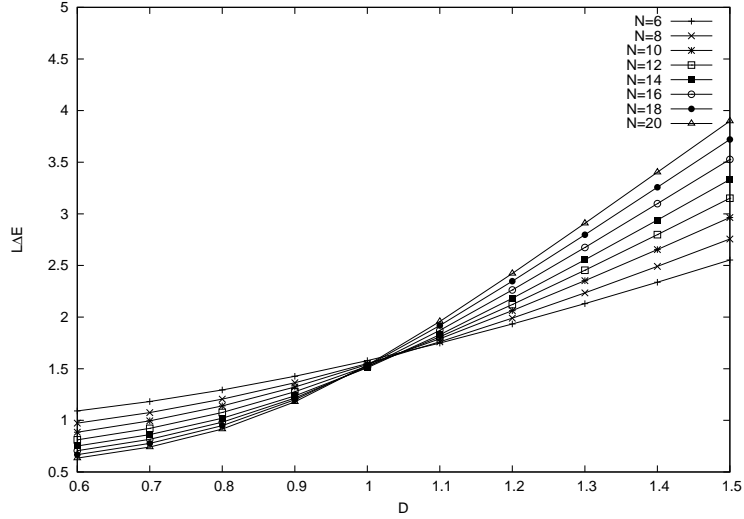}%
\caption{
\label{figure2}
Scaled energy gap 
$L\Delta E$ is plotted for various $D$ and 
$N=6,8,\dots,20$ ($L=\sqrt{N}$);
note that we survey the $D$-driven phase transition
with the other interactions,
$(J_{NN},J_{NNN},D_\Box)$,
adjusted to
a fixed point (\ref{FP_intro}).
We observe a clear indication of the
$D$-driven transition around $D\approx 1$.
Apparently, the finite-size-scaling behavior 
is improved as compared to that of 
the conventional $XY$ model
(Fig. \ref{figure4}).
}
\end{figure}

\begin{figure}
\includegraphics[width=100mm]{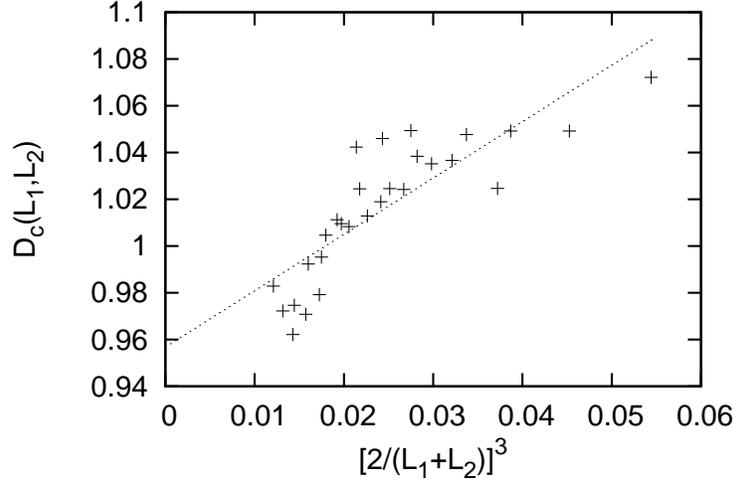}%
\caption{
\label{figure3}
The approximate critical point
$D_c$ (\ref{TP_def}) 
is plotted for $(2/(L_1+L_2))^3$ with 
$6 \le N_1<N_2\le 20$ ($L_{1,2}=\sqrt{N_{1,2}}$).
The least-squares fit to these data yields 
$D_c=0.9569(83)$ 
in the thermodynamic limit $L\to\infty$.
}
\end{figure}

\begin{figure}
\includegraphics[width=100mm]{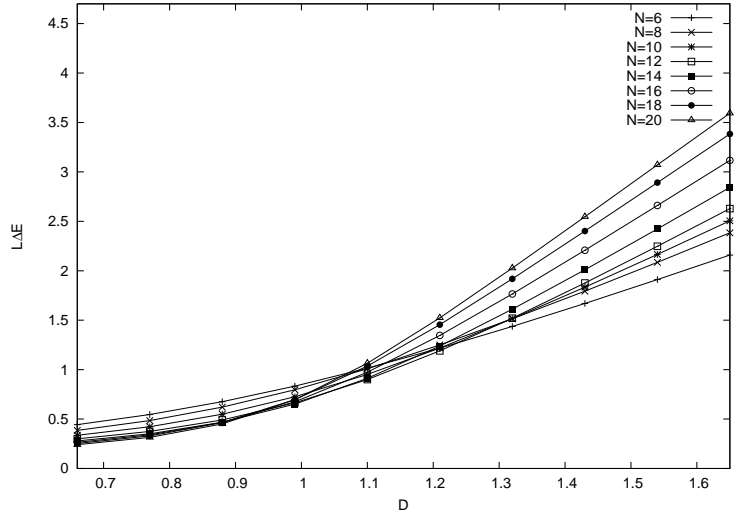}%
\caption{
\label{figure4}
Tentatively,
we turned off the extended interactions,
$(J_{NN},J_{NNN},D_\Box)=(0.2,0,0)$,
and calculated the scaled energy gap
$L\Delta E$
for various $D$ and $N=6,8,\dots,20$ ($L=\sqrt{N}$).
We notice that the data are scattered as compared to those
of Fig. \ref{figure2}.
}
\end{figure}

\begin{figure}
\includegraphics[width=100mm]{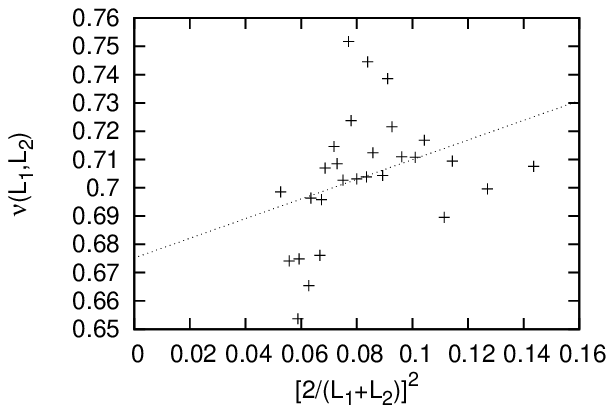}%
\caption{
\label{figure5}
The approximate critical exponent
$\nu$ (\ref{nu_def})
is plotted for $(2/(L_1+L_2))^2$ with 
$6\le N_1<N_2\le 20$ ($L_{1,2}=\sqrt{N_{1,2}}$).
The least-squares fit to these data yields
$\nu=0.675(16)$
in the thermodynamic limit $L\to\infty$.
}
\end{figure}

\begin{figure}
\includegraphics[width=100mm]{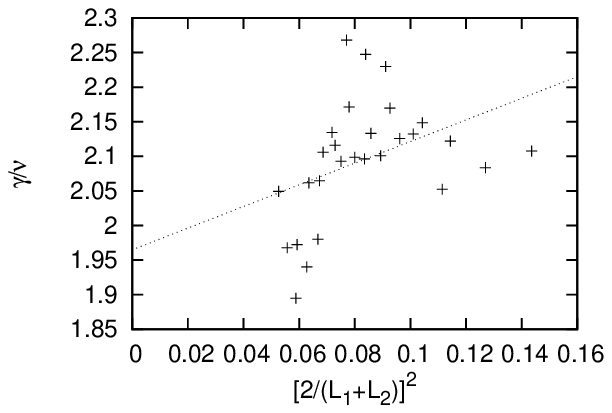}%
\caption{
\label{figure6}
The approximate critical exponent
$\gamma/\nu$ (\ref{gwn_def})
is plotted for $(2/(L_1+L_2))^2$ with 
$6\le N_1<N_2 \le 20$ ($L_{1,2}=\sqrt{N_{1,2}}$).
The least-squares fit to these data yields
$\gamma/\nu=1.965(61)$
in the thermodynamic limit $L\to\infty$.
}
\end{figure}

\begin{figure}
\includegraphics[width=100mm]{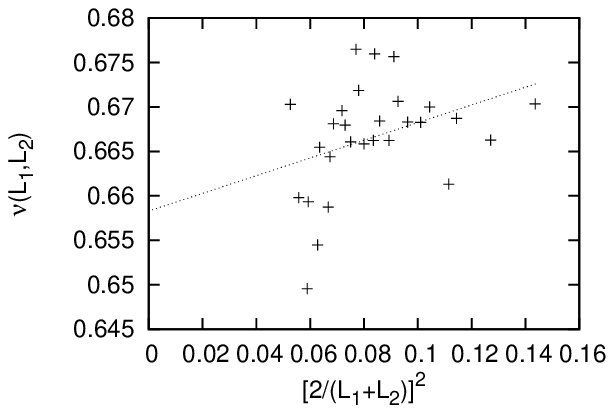}%
\caption{
\label{figure_new1}
The approximate critical exponent
$\nu$ (\ref{nu_def2})
is plotted for $(2/(L_1+L_2))^2$ with 
$6\le N_1<N_2\le 20$ ($L_{1,2}=\sqrt{N_{1,2}}$).
The least-squares fit to these data yields
$\nu=0.658(5)$
in the thermodynamic limit $L\to\infty$.
}
\end{figure}

\begin{figure}
\includegraphics[width=100mm]{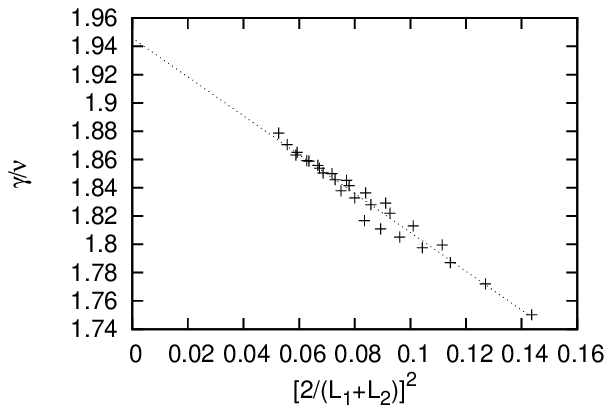}%
\caption{
\label{figure_new2}
The approximate critical exponent
$\gamma/\nu$ (\ref{gwn_def2})
is plotted for $(2/(L_1+L_2))^2$ with 
$6\le N_1<N_2 \le 20$ ($L_{1,2}=\sqrt{N_{1,2}}$).
The least-squares fit to these data yields
$\gamma/\nu=1.946(4)$
in the thermodynamic limit $L\to\infty$.
}
\end{figure}

\begin{figure}
\includegraphics[width=100mm]{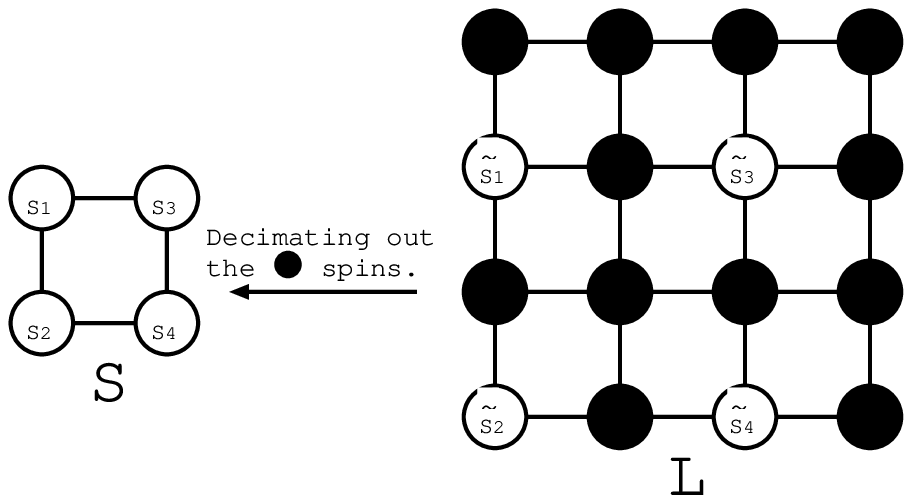}%
\caption{
\label{figure7}
A schematic drawing of the real-space 
renormalization group (decimation)
for the $d=2$ $XY$ model (\ref{Hamiltonian})
is presented.
Through decimating out the spin variables indicated by the symbol $\bullet$
within the $L$ cluster,
we obtain a coarse-grained lattice identical to the
$S$ cluster.
Imposing the scale-invariance conditions, Eqs. (\ref{SC1})-(\ref{SC3}),
we arrive at the fixed point (\ref{FP_intro}).
}
\end{figure}

Last, we argue the validity of the above solution
(\ref{FP_app}) and the boundary condition $b=0.7$.
In Sec. \ref{section3}, 
via the finite-size-scaling analysis,
we obtained $D_c=0.957(25)$
(\ref{transition_point}).  %% with the coupling constants,
%%$(J_{NN},J_{NNN},D_\Box)$, set to Eq. (\ref{FP_app}).
Apparently, this result is
consistent with 
$D=1$ postulated in Eq. (\ref{Dwo1ni}).
Moreover, the simulation data in Fig. \ref{figure2} exhibit
suppressed finite-size corrections, as compared to those of the ordinary $XY$ model,
Fig. \ref{figure6}.
These features
validate the choice of the boundary condition $b=0.7$
as well as the reliability of the fixed point (\ref{FP_app}).
Furthermore, we point out
that the boundary condition $b=0.7$ is reminiscent of
$b=0.4$ utilized in the fixed-point analysis
of the $d=3$ Ising ferromagnet \cite{Nishiyama06}.


%\begin{figure}
%\includegraphics[width=100mm]{figure8.eps}%
%\caption{
%Boundary interaction
%\label{figure7}
%}
%\end{figure}


\begin{thebibliography}{99}

%%\input{bib.tex}

\bibitem{Novotny90}M. A. Novotny, J. Appl. Phys. {\bf 67}, 5448 (1990).
\bibitem{Novotny92}M. A. Novotny, Phys. Rev. B {\bf 46}, 2939 (1992).
\bibitem{Novotny91}M. A. Novotny, in
{\it Computer Simulation Studies in Condensed Matter Physics III},
edited by D. P. Landau, K. K. Mon, and H.-B. Sch\"uttler (Springer-Verlag, Berlin, 1991).

\bibitem{Nishiyama04}Y. Nishiyama, Phys. Rev. E {\bf 70}, 026120 (2004).
\bibitem{Nishiyama07}Y. Nishiyama, Phys. Rev. E {\bf 75}, 011106 (2007).


% experiment
\bibitem{Barmatz07}
M. Barmatz, I. Hahn, J. A. Lipa, and R. V. Duncan,
Rev. Mod. Phys. {\bf 79}, 1 (2007).
%highT
\bibitem{Buteral97} %%  2-d nu?? amplitude??
P. Butera and M. Comi,
%%Phys. Rev. B, {\bf 60}, 6749 (1999).
% nu gamma
Phys. Rev. B {\bf 56}, 8212 (1997).
% field theory 
\bibitem{Guida98}
R. Guida and J. Zinn-Justin,
J. Phys. A {\bf 31}, 8103 (1998).
%B 21 3976 80  
%variational PE
\bibitem{Kleinert05}
H. Kleinert and V. I. Yukalov,
Phys. Rev. E {\bf 71}, 026131 (2005).
%MC   0.6717(3)
\bibitem{Burovski06}
E. Burovski, J. Machta, N. Prokof'ev, and B. Svistunov,
Phys. Rev. B {\bf 74}, 132502 (2006).
% MC+HighT a=-0.0151(3) n=0.6717(1) eta=0.0381(2) gamma=1.3178(2) be=0.3486(1) del=4.780(1)
\bibitem{Campostrini06}
M. Campostrini, M. Hasenbusch, A. Pelissetto, and E. Vicari,
Phys. Rev. B {\bf 74}, 144506 (2006).


% S=1 
\bibitem{Costa03}
B.V. Costa and A.S.T. Pires,
J. Mag. Mag. Mat. {\bf 262}, 316 (2003).
\bibitem{Leonel06}
S.A. Leonel, A.C. Oliveira, B.V. Costa, and P.Z. Coura,
J. Mag. Mag. Mat. {\bf 305}, 157 (2006).
\bibitem{Dely06}
J. Dely, J. Stre\v{c}ka, and L. \v{C}anov\'{a},
cond-mat/0611212.


%
% elimination
\bibitem{Blote96}
H. W. J. Bl\"ote, J. R. Heringa, A. Hoogland, E. W. Meyer,
and T. S. Smit,
Phys. Rev. Lett. {\bf 76}, 2613 (1996).
\bibitem{Nishiyama06}
Y. Nishiyama,
Phys. Rev. E {\bf 74}, 016120 (2006).
% perfect action
\bibitem{Ballesteros98}
H.G. Ballesteros, L.A. Fern\'andez, V. Mart\'in-Mayor, and
A. Mu\~noz Sudupe,
Phys. Lett. B {\bf 441}, 330 (1998).
% Impoved
\bibitem{Hasenbusch99}
M. Hasenbusch and T. T\"or\"ok,
J. Phys. A {\bf 32}, 6361 (1999).



% large D
\bibitem{Botet83}
R. Botet, R. Jullien, and M. Kolb,
Phys. Rev. B {\bf 28}, 3914 (1983).


%
\bibitem{Gagliano87}
E. R. Gagliano and C. A. Balseiro,
Phys. Rev. Lett. {\bf 59}, 2999 (1987).


%
\bibitem{Nishiyama07b}
Y. Nishiyama,
Phys. Rev. E {\bf 75}, 051116 (2007).



\bibitem{Lipa03}
J.A. Lipa, J.A. Nissen, D.A. Stricker, D.R. Swanson, and T.C.P. Chui,
Phys. Rev. B {\bf 68}, 174518 (2003).
% disagree
\bibitem{Pogorelov07}
A. A. Pogorelov and I. M. Suslov,
JETP Lett. {\bf 86}, 39 (2007).

%
\bibitem{Itakura03}
M. Itakura,
J. Phys. Soc. Jpn. {\bf 72}, 74 (2003).

%
\bibitem{Swendsen82}
R. H. Swendsen, in
{\it Real-Space Renormalization},
edited by T. W. Burkhardt and J. M. J. van Leeuwen
(Springer-Verlag, Berlin, 1982). 




\end{thebibliography}
\end{document}